\documentstyle[12pt]{article}

\input{tcilatex}

\begin{document}

\title{ESTIMATE\ ON \ DECELERATION PARAMETER IN A UNIVERSE WITH VARIABLE FINE
STRUCTURE ``CONSTANT \ ''.}
\author{Marcelo S.Berman$^{(1)}$ \and and Luis A.Trevisan$^{(2)}$ \\
(1) Tecpar-Instituto de Tecnologia do Paran\'{a}\\
Grupo de Projetos Especiais.\\
R.Prof Algacir M. Mader 3775-CEP 81350-010\\
Curitiba-PR-Brasil\\
Email: marsambe@tecpar.br\\
(2) Universidade Estadual de Ponta Grossa,\\
Demat, CEP 84010-330, Ponta Grossa,Pr,\\
Brasil \ email: latrevis@uepg.br}
\maketitle

\begin{abstract}
We determine a cosmological model \ that \ includes acceleration of the
Universe \ and \ time-decreasing fine structure ``constant '', as found
recently. We found that the present -day deceleration \ parameter should be $%
q\simeq -1.1\times 10^{-5}$. The \ variation in \ $\alpha $ \ is \
attributed to a slowly increasing \ speed of light.

Pacs 98.80 Hw
\end{abstract}

\noindent \newpage

\ 

\begin{center}
\bigskip ESTIMATE ON DECELERATION PARAMETER IN A UNIVERSE WITH VARIABLE FINE
STRUCTURE ``CONSTANT ''.

\bigskip

Marcelo S. Berman and Luis A.\ Trevisan
\end{center}

The \ latest developments of physics, \ when we enter this century, are the
discovery that the Universe is accelerating, and that the fine structure
``constant '' is \ varying with the age of the Universe.

The first conclusion, based on Supernovae \cite{[1]} observations may mean
that the value of the cosmological constant is non-null; however, what we
know for sure is that that the deceleration parameter,

\begin{equation}
q=-\frac{\ddot{R} R}{\dot{R}^{2}}
\end{equation}
is negative. In (1), $R=R(t)$ stands for the scale-factor, and overdots are
time-derivatives, while the metric is Robertson-Walker's.

The second finding is that the ``constant ''

\begin{equation}
\alpha \equiv \frac{e^{2}}{\frac{h}{2\pi }c},
\end{equation}
is varying  with time.

Let us call \ $F=\Delta \alpha /(\alpha \Delta t)$ (experimental). We can
make a rough estimate for $F$ by employing \ the \ values found by \ Webb et
\ al for \ the \ deviation from the average:

\begin{equation}
\frac{\Delta \alpha }{\alpha }\simeq -0.72\times 10^{-5}
\end{equation}
spanning $23\%$ \ to \ $87\%$ of the age of the Universe \cite{[2]}. A \
rough estimate for \ $\Delta t$ is given by \ $\Delta t=(0.87-0.23)t$ where $%
t$ stands for the age of \ the Universe. Any competent \ reader \ will be
able to \ modify our \ results when a better estimate shall be available in
the literature for \ $F.$

As in cosmology, the charge of the electron, or Planck's constant, do not
play a large-scale r\^{o}le, we fix our attention on a time-varying speed of
light, $c=c(t).$

Webb et al.\cite{[2]} pointed out that a common property of unified
theories, applied to Cosmology, is that they predict space and
time-dependence of the coupling constants. Let us suppose that \ Newton's
gravitational ``constant'' $G$ \ is also time varying; the natural framework
we need is JBD ( Jordan-Brans-Dicke) theory with varying speed, as Barrow
has been studying \cite{[3]}.

It is well \ accepted that, for the present Universe a power-law of time can
represent the scale-factor. We are thus led to consider constant
deceleration parameter models, as studied initially \ by Berman \cite{[4]},
and Berman and Gomide\cite{[5]}. On defining $\ m=1+q=$constant, we find,

\begin{equation}
R=R(t)=(mDt)^{\frac{1}{m}}
\end{equation}
where $D=$constant.

Einstein-de Sitter model, with $k=0$ ($k$ is the tricurvature), is of this
kind, with $m=3/2.$ As our time scale for the present Universe duration so
far is about 10$^{10}$ years, and there is not evidence for a significant
time variation of $q,$ for the present Universe, we find that a constant
deceleration parameter model is reasonable for the present Universe. In
reference [7] there is a comment on one single evidence for a changing $q$
for the Universe in its present phase, but obviously this is not a
conclusive evidence up to now. We shall show that the solution with variable 
$\alpha $ differs from Einstein-de Sitter's model.

Barrow's equation read:

\begin{equation}
H^{2}=\frac{8\pi \rho }{\phi }-\frac{\dot{\phi}}{\phi }H+\frac{\omega }{6}%
\frac{\dot{\phi}^2}{\phi^2}-\frac{kc^{2}(t)}{R^{2}}
\end{equation}

\begin{equation}
\frac{3\ddot{R}}{R}=\frac{-8\pi }{(3+2\omega )\phi }\left[ (2+\omega )\rho +%
\frac{3(1+\omega )p}{c^{2}(t)}\right] -\frac{\ddot{\phi}}{\phi }-\frac{%
\omega \dot{\phi}^{2}}{\phi ^{2}}
\end{equation}

\begin{equation}
\ddot{\phi}+3H\dot{\phi}=\frac{8\pi }{3+2\omega }\left( \rho -\frac{3p}{%
c^{2}(t)}\right)
\end{equation}

Unlike Barrow, we shall not impose a definite equation of state; the values
for energy density $\rho $ and cosmic-pressure $p$ will arise naturally in
our solution.

From the theory of constant deceleration parameter we have:

\begin{equation}
H=(mt)^{-1}
\end{equation}
We find the following solution:

\begin{equation}
R=(mDt)^{\frac{1}{m}}
\end{equation}

\begin{equation}
\phi =At^{n}
\end{equation}

\begin{equation}
\rho =ABt^{n-2}
\end{equation}

\begin{equation}
c=Ct^{\frac{1}{m}-1}
\end{equation}

\begin{equation}
p=Ft^{\gamma -4+2/m}
\end{equation}
where $A,B,\gamma ,F,C$ are constants, related by means of equations (4),
(5) and (6).

From (11) and (2), we find,

\begin{equation}
\frac{\dot{\alpha}}{\alpha }=Hq
\end{equation}
Let us now compare with the experimental values:

\begin{equation}
\frac{\Delta \alpha }{\alpha \Delta t}=\frac{\Delta \alpha }{\alpha \left[
0.64\frac{H^{-1}}{m}\right] }\simeq -\frac{0.72}{0.64}10^{-5}Hm
\end{equation}
or \ 

\begin{equation}
\frac{\Delta \alpha }{\alpha \Delta t}\simeq 1.1\times 10^{-5}Hm.
\end{equation}
For \ the age of the Universe, we make use of relation (7). We see that $%
q\simeq -1.1\times 10^{-5\text{ }}$ and $\ m\simeq 0.99999.$

The speed of light is, then increasing slowly with time. The Universe has a
minimal, though finite, acceleration, as can be checked from (8), with the
values of $q$ and $m.$

It has been a happy coincidence, that when the $\alpha $ varying experiments
were published, we were able to apply the recent Barrow-Jordan-Brans-Dicke
equations, in order to derive a value for $q,$ which was lacking in the
literature. Barrow and Magueijo [6] explained how a changing $\alpha $ could
yield the Supernova results with a different argumentation than ours. For up
to date references in relation to variable $G$ cosmologies, see ref [8] [9].
We refer to these papers for further information.

\bigskip {\bf Acknowledgments}

Both authors thank support by Prof Ramiro Wahrhftifig, Secretary of Science
and Technology of the State of Parana, and from our Institutions, especially
to Jorge\ L.Valgas, Roberto Merhy, Mauro K. Nagashima, Carlos Fior, C.R.
Kloss, J.L.Buso, and Roberto Almeida. M.S.B. thanks also his two intelectual
mentors, Prof. M.M.Som and Prof. F.M. Gomide.

\end{document}